# Bistable all-optical devices based on nonlinear epsilon-near-zero (ENZ) materials.

J. Gosciniak[1*], Z. Hu[2], M. Thomaschewski[2], V. Sorger[2,3], and J.B. Khurgin[4^]

[1]ENSEMBLE3 sp. z o.o., Wolczynska 133, 01-919 Warsaw, Poland
[2]Department of Electrical and Computer Engineering, George Washington University, Washington DC, 20052, USA.
[3]Optelligence LLC, 10703 Marlboro Pike, Upper Marlboro, 20772, MD, USA
[4]Electrical and Computer Engineering Department, Johns Hopkins University, Baltimore, Maryland 21218, USA
Corresponding authors: * jeckug10@yahoo.com.sg, ^ jakek@jhu.edu

**Abstract**

Non-linear and bistable optical systems are a key enabling technology for the next generation optical networks and photonic neural systems with many potential applications in optical logic and information processing. Here, we propose a novel bistable resonator-free all-optical waveguide device based on indium tin oxide as nonlinear epsilon-near-zero material providing a cost-efficient and high-performance binarity photonic platform. The salient features of the proposed device are compatibility with silicon photonics, enabling sub-picosecond operation speeds with moderate switching power. The device can act as an optical analogue of memristor or thyristor and can become an enabling element of photonic neural networks not requiring OEO conversions.

**Introduction**

Bistability is a property of a system that exhibits two stables steady states and the system rests in one of those states depending on the history of this system. For example, a permanent magnet can be in one of two states with oppositely directed magnetization [1], depending on the sign of the magnetic field that had been applied and then removed. Two states thus can represent two values of a binary digit (bit) in magnetic storage. Similarly, resistance of many electronic devices, such as thyristors [2] and memristors [3] can be low or high depending on whether a high current had or had not been applied. Bistability is the underlying principle for switching logic and storage applications. While electronic and magnetic bistable devices of various types are in wide use, when it comes to applications in the optical domain [4-6], the situation is less sanguine as no bistable elements with acceptable characteristics (fast speed, low power, wide optical bandwidth) have emerged, and not for the lack of trying. Most of the proposed and demonstrated bistable optical devices are based on a combination of a nonlinear optical material and a resonant cavity, require high optical powers for switching and are often incompatible with integrated photonics [7-9]. A different type of bistable device is a electro-absorptive self-electro-optic device (SEED) [10], constructed of a semiconductor multiple quantum wells biased by an external voltage, which creates an bistable shifts of the absorbing wavelengths of the transmitted light and does not require optical cavity. However, the device requires electrical bias and hence cannot be considered all-optical, and its switching time is limited by the incorporated RC constant of the electrical circuit [11]. All the devices operate within a relatively narrow optical bandwidth determined by either the resonance wavelength in nonlinear cavity-based systems or the excitonic electro-absorption bandwidth in in SEEDs. The lack of reliable bistable all-optical devices impedes technological progress in many fields of conventional and quantum optics [12], hindering particularly the development of all-optical neural networks that exhibit bistable dynamics [13-15].

**New flavor of bistability with ENZ waveguides**

Here we propose a fundamentally new concept of all-optical bistable devices that do not require a resonator, enabling thereby broadband operation, which can be integrated into a photonic waveguide and operate at sub-ps switching speeds. The proposed device is based on the unique properties of epsilon-near-zero (ENZ) materials (typically transparent conductive oxides (TCO) like indium tin oxide (ITO) [16-18]), which exhibits a dispersion of its relative electrical permittivity that combines responses of bound ($\varepsilon_\infty$) and free electrons according to

$$\varepsilon(\omega) = \varepsilon_\infty - \frac{\omega_p^2}{\omega^2 + i\omega\gamma} \qquad (1)$$

where $\varepsilon_\infty$ is permittivity due to the bound electrons, $\omega_p=(Ne^2/\varepsilon_0 m^*)^{1/2}$ is plasma frequency, $N$ is the carrier density, $m^*(E)$ is the energy-dependent effective mass, and $\gamma$ is the scattering rate. The real part of permittivity changes from positive to negative at the so-called ENZ frequency $\omega_0=\omega_p/\varepsilon_\infty^{1/2}$ (identified as screened plasma frequency) with an imaginary part of the permittivity introducing absorption in the material, as shown in Fig. 1a and b. Due to the large permittivity tunability, transparent conductive oxides have been used extensively for electro-optic modulation [19, 20] and other optoelectronic applications (e.g., in display technology or vertical-cavity surface-emitting lasers) [21-23]. However, its versatility extends beyond existing applications.

Consider the interface of ENZ material in the vicinity of $\omega_0$ with $Re(\varepsilon(\omega_0))\approx0$ and a conventional dielectric with $\varepsilon_d>0$.

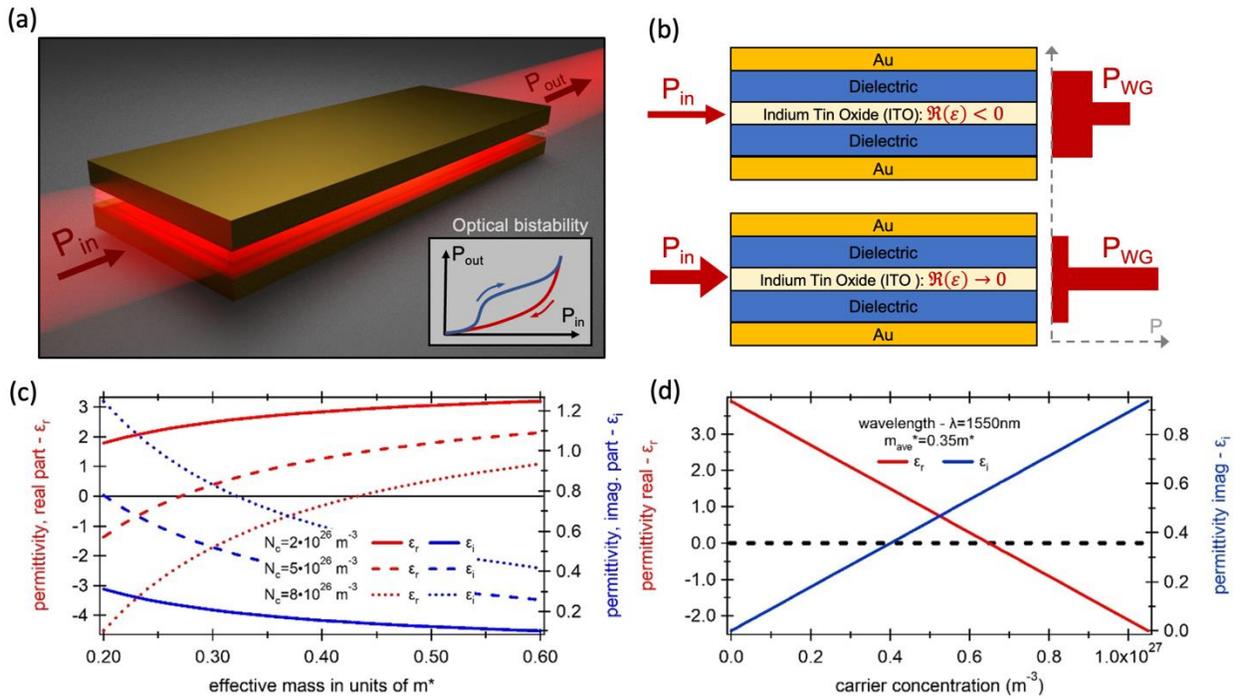

**Fig.1 (a)** Working principle of the proposed bistable all-optical device based on an epsilon-near-zero material (indium tin oxide, ITO). **(b)** At low optical powers (top) the ITO layer is metallic and the optical power density in it is low. As the power increases (bottom) the real part of permittivity approaches zero and the optical power inside multilayered system gets redistributed which shifts *Re(ε)* even closer to zero, thus closing the positive feedback loop and causing switching. **(c, d)** Dispersion of real and imaginary parts of dielectric permittivity as a function of (c) effective mass and (d) carrier concentration.

According to boundary conditions for light polarized normal to interface (TM), the displacement *D=εE* is continuous across the interface, and therefore the electric field (and hence the energy density) is much higher in the ENZ material than in the conventional dielectric. If a thin layer of ENZ material is inserted in the slot of a waveguide (which can be photonic or plasmonic) as shown in Fig. 1b the closer is the wavelength to $\omega_0$, the higher is the field the slot. This is a well-studied slot waveguide effect [24, 25] that allows subwavelength field confinement in the low permittivity medium, taken to its ultimate ε~0 limit.

Unlike the dielectric cladding, ENZ material is highly absorptive and as the optical field concentrated inside the slot increases, the absorption increases. At the point of large electric fields, the nonlinear properties of the TCO-based ENZ materials come into play [16, 26, 27]. When light is absorbed by the TCO material, it causes an increase of the electron temperature in the conduction band, that in turn leads to an increase of the effective mass and corresponding rise of the dielectric constant, according to Eq. (1). If the carrier density is such that in the absence of light the real permittivity is *Re(ε)*<0, an increase of *Re(ε)* would move it closer to zero, which in turn will cause even higher concentration of energy inside the ENZ material layer. Thus, a positive feedback mechanism, essential for bistability, is established. As more energy is concentrated in the ENZ layer, the electron temperature in it rises causing an increase of the dielectric constant and further concentration of energy in the ENZ layer. Obviously, no stable solution can exist with positive feedback and switching occurs.

**Mechanism of bistability and switching**

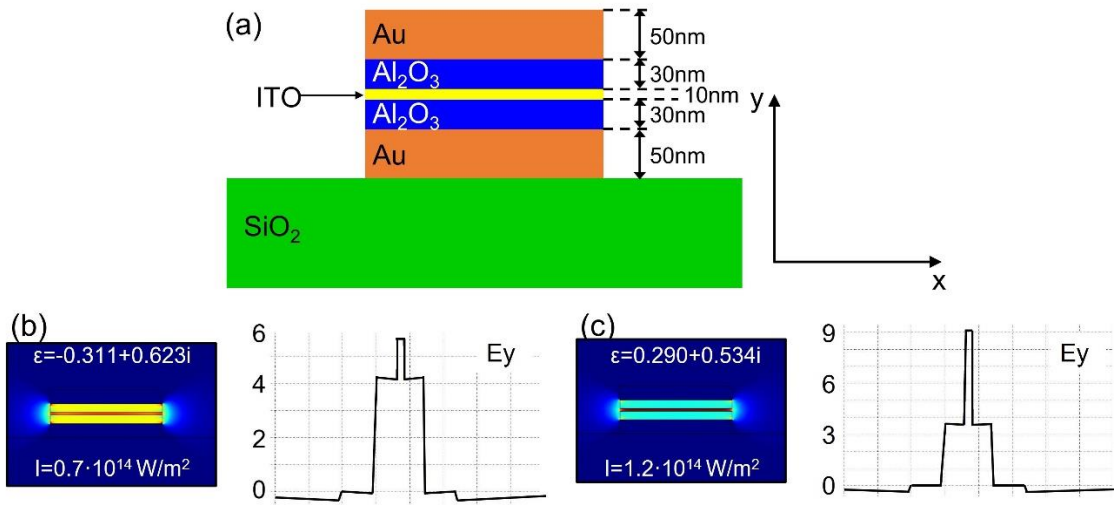

**Fig.2** (a) Geometry of the proposed bistable device and (b, c) electric field distribution in the state of (b) high and (c) low absorption regime.

The proposed concept of bistability is investigated using two-dimensional finite element method (FEM) simulations at the telecom wavelength of 1550 nm using commercial software COMSOL.

The investigated arrangement consists of a typical metal-insulator-metal (MIM) structure where $Al_2O_3$ serves as an insulator and the ENZ material is placed in the center of the $Al_2O_3$ (Fig. 1). Here, the ITO was used as the ENZ material as it is the most widely used TCO material in industry with very well-established fabrication process that shows the presence of the ENZ region at tolerable carrier concentrations [22, 24]. For the MIM plasmonic structure, the electric field of the propagating mode exhibits maxima at both metal-insulator interfaces and decays exponentially into the insulator (Fig. 2b and c). The relative amount of energy in the metal and the insulator depends on the optical properties of the material and the waveguide geometry.

The thickness of Au was chosen to be $h$=50 nm, while the thickness of $Al_2O_3$ and ITO is $h_1$=30 nm and $h_2$=10 nm, respectively. The width of the multilayer stack is $w$=400 nm. For the calculations, the complex refractive index of Au is assumed to be $n_{Au}$=0.596+10.923$i$ [26] while for the refractive index of $Al_2O_3$ is $n_{Al2O3}$=1.621+0.00008$i$ [27]. The calculations are performed for a thermalization time $\tau$=100 fs, the Fermi energy $E_F$=4.5 eV, and the carrier concentration $N_c$=8·$10^{26}$ $m^{-3}$.

Let us now quantify the switching and bistability in a specific waveguide shown in Fig.2. We model the waveguide absorption coefficient α of light at a given wavelength $\lambda$=1500 nm, $\omega$=2π×200 THz as a function of $\omega_p^2$. As shown in Fig. 3a, the maximum of absorption corresponds to $\omega_0=\omega_p/\varepsilon_\infty$=2π×200 THz, i.e., to the minimum of $Re(\varepsilon)$. It is obviously the expected consequence of an electric field concentrated in the slot where absorption takes place. One can then estimate the power loss per unit volume of ITO. For an optical power $P$ in the waveguide, the absorbed power density in the ITO filled slot is $P_{abs}=\alpha P/S_{ITO}$, where $S_{ITO}$ is the cross-section of the ITO layer. The density of energy transferred to the hot carriers in the ITO is $U_{hot}=P_{abs}\tau=\alpha P\tau/S_{ITO}$, where $\tau$ is the "cooling time", i.e., the time it takes to transfer energy from the hot carriers to the lattice *via* electron-phonon scattering. This time is typically a few hundreds of femtoseconds [28]. As shown in [26, 27] when the hot carriers gain energy, they move higher into the conduction band of ITO where their effective mass is increased which introduces a change of $\omega_p$ impacting the effective mass as according to [26]

$$\frac{\delta m^*}{m^*} = -\frac{\delta \omega_p^2}{\omega_p^2} \approx \frac{U_{hot}}{N\hbar k_F v_F}, \quad (2)$$

Where $k_F$ and $v_F$ are respectively Fermi wave vector and velocity. The relationship to the Fermi energy for a parabolic band follows $\hbar k_F v_F=2E_F$ and for an extremely non-parabolic band (such as graphene) follows $\hbar k_F v_F=E_F$, where $E_F$ is Fermi energy. For the sake of simplification, we assume that $\hbar k_F v_F \approx E_F$ which is physically valid – in order to change the effective mass $U_{hot}$ of each electron in an assembly of $N$ electrons, the change of energy of each electron $U_{hot}/N$ should be noticeable relative to the energy it already had, i.e., roughly Fermi energy. We then obtain the following linearized relation between the absorption coefficient, plasma frequency and power

$$\omega_p^2 = \omega_{p0}^2 - \omega_{p0}^2 \frac{\alpha \tau}{NE_F S} P, \quad (3)$$

Where $\omega_{p0}$ is the plasma frequency in the absence of optical power and heating. In our example the carrier density is 8·$10^{26}$ $m^{-3}$ and $E_F$=4.5 eV. We can now plot a family of straight lines (Eq. 3) for different values of power $P$ in the Fig. 3a. Depending on the value of $P$ the curve $\alpha(\omega_p^2)$ shows either one or three intersections with straight lines. In the case of three intersection points (on the right slope of the curve) the middle solution is unstable due to positive feedback. Therefore, gradually increasing power (right to left) leads to gradual increasing both effective mass and loss until the critical (switching) value $P_{up}$. When the power line gets tangential to the curve $\alpha(\omega_p^2)$ the positive feedback generates a stable solution to switch at a much higher value of the effective mass and loss. During switching, $Re(\varepsilon)$ changes the sign from negative to positive and a further increase in power no longer causes an increase of the degree of carrier concentration in the ITO which induces a stable but gradually changing solution on the left slope of the curve. When the power is cycled down the high-loss regime, the solution remains stable until the second switching value $P_{down}<P_{up}$ is reached, and the positive feedback enables switching into the low-loss state with $Re(\varepsilon)<0$. Thus, a hysteresis region is established as shown in Fig. 3b showing the absorptive loss per unit length as a function of the propagating power.

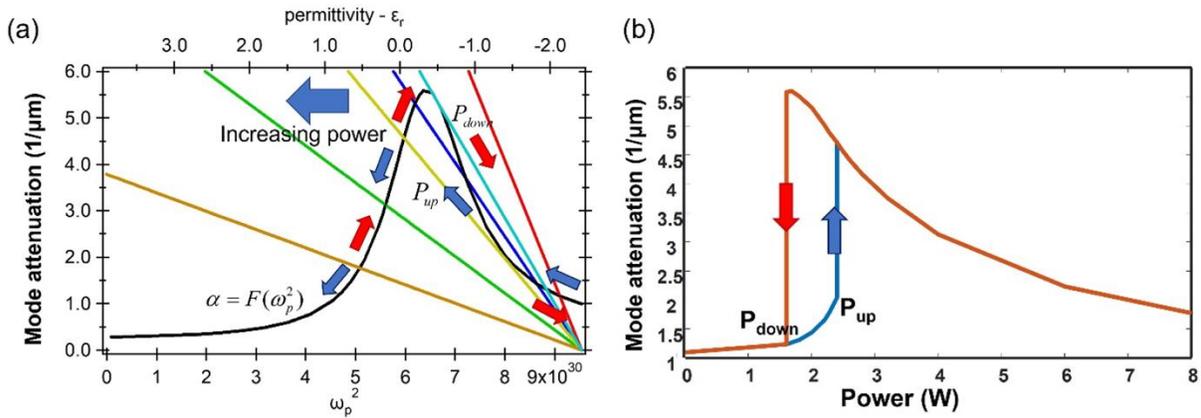

**Fig3 (a)** Illustration of bistability and switching as power changes in the waveguide. (b) Absorptive loss as a function of the propagating power exhibiting hysteresis manifesting all-optical bistability.

**Evaluating the performance of bistable switch**

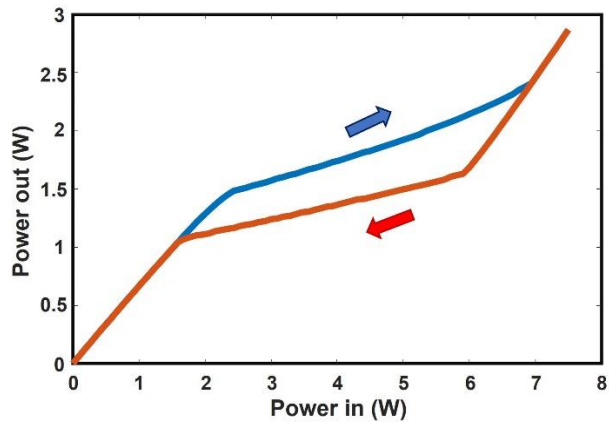

**Fig.4** Input-output characteristics of the proposed bistable device of 1 micron length

The hysteresis curve of Fig. 3b is a local curve, i.e., which relates to the absorption $\alpha(z)$ and propagating power $P(z)$ at a given point $z$. To find a relation between the input and output power at $z=L$, we need to integrate the coupled system of equations

$$\alpha(z) = \alpha(P(z))$$
$$\frac{dP(z)}{dz} = -\alpha(z)P(z) \tag{4}$$

with boundary condition $P(0)=P_{in}$. The results are shown in Fig. 4 for the case of $L$=1 µm. The switching does not occur simultaneously over the entire length of the device but propagates shock-wave-like from the front as local $P(z)$ reaches the switching power $P_{up}$. As input power is ramped down, the "switching wave" propagates backward. Thus, the hysteresis curve has a tilted shape (Fig. 4). Presence of hysteresis indicates that the proposed nonlinear device can serve as an optical analogue of memristor. Alternatively, if a separate switching pulse is coupled to the device, then it becomes optical analogue of thyristor.

The bistable characteristics of the proposed device providing fast and reliable all-optical binarity can have impactful applications in artificial neural networks with binary weights and nonlinear activations. Binarization in neural networks enjoy several hardware-friendly properties including memory saving, power efficiency and significant acceleration. For instance, computationally heavy matrix multiplication operations can be replaced with light weighted bitwise XNOR operations and Bitcount operations. Furthermore, it has been shown that binary neural networks significantly improve the robustness of the network by keeping the magnitude of the noise small [29], which can be integrated to an optical neural network with our proposed device providing low power consumption, high computational speed and large information storage.

**Bistable ENZ device in Neural Networks**

All-optical neural networks have been developed in recent years, and this field requires all-optical nonlinear activation functions. Many researchers are trying to develop nonlinear activation functions [14, 30-32]. Bistable devices can be used as nonlinear activation functions of all-optical neural networks. Such bistable devices can be easily implemented into integrated optical deep neural networks [29-36] since they are all waveguide-based systems. It can also be implemented into free-space optical neural networks [37, 38] with pixel-wise modulation array as SLM. Also, if we need to deal with time series data like NLP tasks, we can match the system clock to the response time of the bistable material, we will take advantage of the temporal dynamic behavior of the bistable material, the activation function output depends on the previous state.

To explore the applications of the proposed bistable ENZ device in the all-optical neural networks, we simulate fully connected deep neural networks using our bistable characteristics as the activation function. In the simulation, we use MNIST, FashionMNIST and CIFAR-10 as test datasets. The neural network consists of flat images as input layer, 2 hidden layers with 400 and 200 nodes, and 10 output nodes for classification (Fig. 5). Simulations show that our bistable nonlinear activation function performs similarly to the ReLu and Sigmoid, the classic digital nonlinear activation functions, and they all show a prediction accuracy of 98 % for MNIST, 89 % for FashionMNIST, and approximately 52 % for CIFAR-10. Compared to a neural network without a nonlinear activation function, MNIST shows only 92 %, FashionMNIST is 84 % and CIFAR-10 is 38 %. (https://github.com/Sorger-Lab/General/tree/master/Bistable)

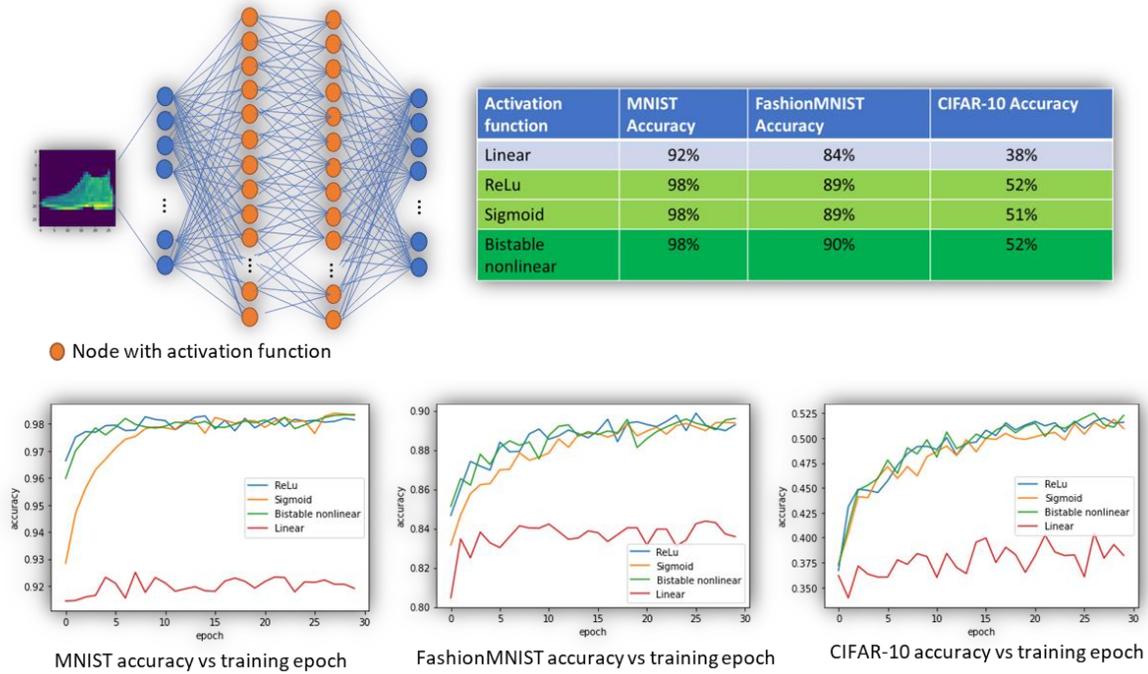

**Fig. 5** The Neural Network performance with Bistable nonlinear activation function. We simulate a neural network with two hidden layers with four different activation functions, no activation, ReLu, Sigmoid, and Bistable nonlinear. The simulation shows our Bistable nonlinear has similar performance with classic digital nonlinear activation functions ReLu and Sigmoid, 98 % for MNIST dataset, ~90 % for FashionMNIST dataset, and 52 % for CIFAR-10 dataset comparing to the linear activation function with 92 %, 84 %, and 38 %.

**Conclusions**

In this work we have proposed and analyzed a novel bistable device based on ENZ material in a slot waveguide and have shown that it shows high nonlinearity and significant bistability without optical cavity. Small footprint and high switching speed of the proposed device can lead to applications in various all-optical systems including neural networks where ethe performance of the devices was estimated and found to be excellent.

**Acknowledgments**

J.G. thanks the "ENSEMBLE3 - Centre of Excellence for Nanophotonics, advanced materials and novel crystal growth-based technologies" project (GA No. MAB/2020/14) carried out within the International Research Agendas program of the Foundation for Polish Science co-financed by the European Union under the European Regional Development Fund and the European Union's Horizon 2020 research and innovation program. Teaming for Excellence (GA. No. 857543) for support of this work.  J.K would like to thank DARPA NLM program.